\documentclass[11pt]{article}
\usepackage{moriond,epsfig,amsmath,amssymb}

\def\be{\begin{equation}}
\def\ee{\end{equation}}
\def\bea{\begin{eqnarray}}
\def\eea{\end{eqnarray}}

\newcommand{\Dmq}{\Delta m^2}

\newcommand{\stheta}{\sin^22\theta_{13}}

\begin{document}
\vspace*{4cm}
\title{THE LSND PUZZLE IN THE LIGHT OF MINIBOONE RESULTS~\footnote{Talk given at Rencontres de Moriond EW 2008, La Thuile, 1--8 March 2008}}

\author{THOMAS SCHWETZ}

\address{Physics Department, Theory Division, CERN, CH--1211
  Geneva 23, Switzerland}

\maketitle\abstracts{I give a brief overview over various attempts to
reconcile the LSND evidence for oscillations with all other global
neutrino data, including the results from MiniBooNE. I discuss
the status of oscillation schemes with one or more sterile neutrinos
and comment on various exotic proposals.}

\section{Introduction}
 
\begin{figure}
  \centering
  \includegraphics[height=6cm]{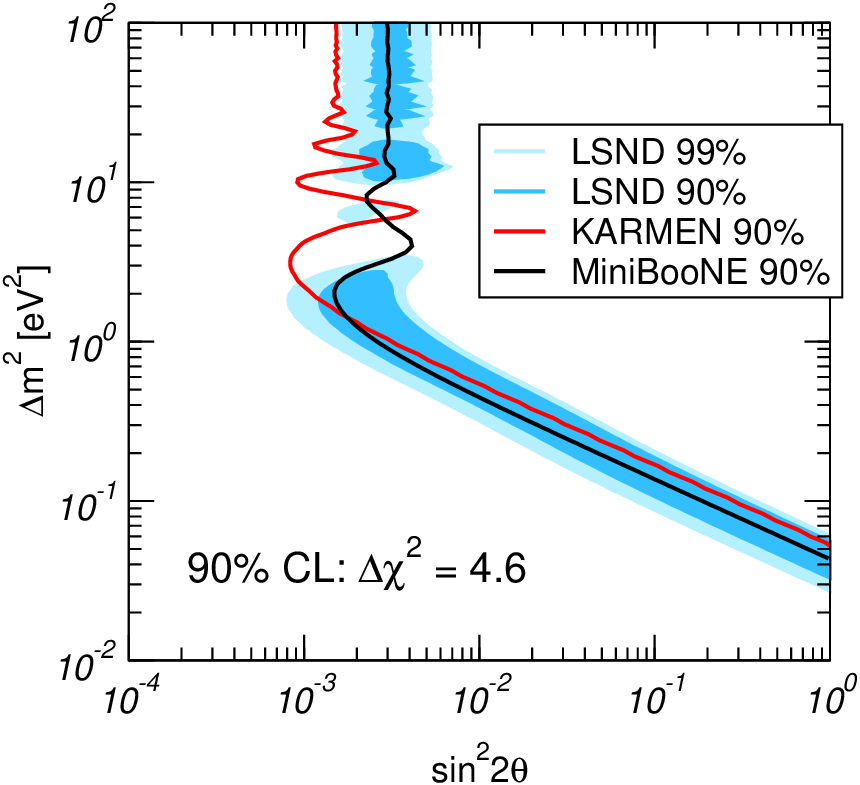}\quad
  \includegraphics[height=6cm]{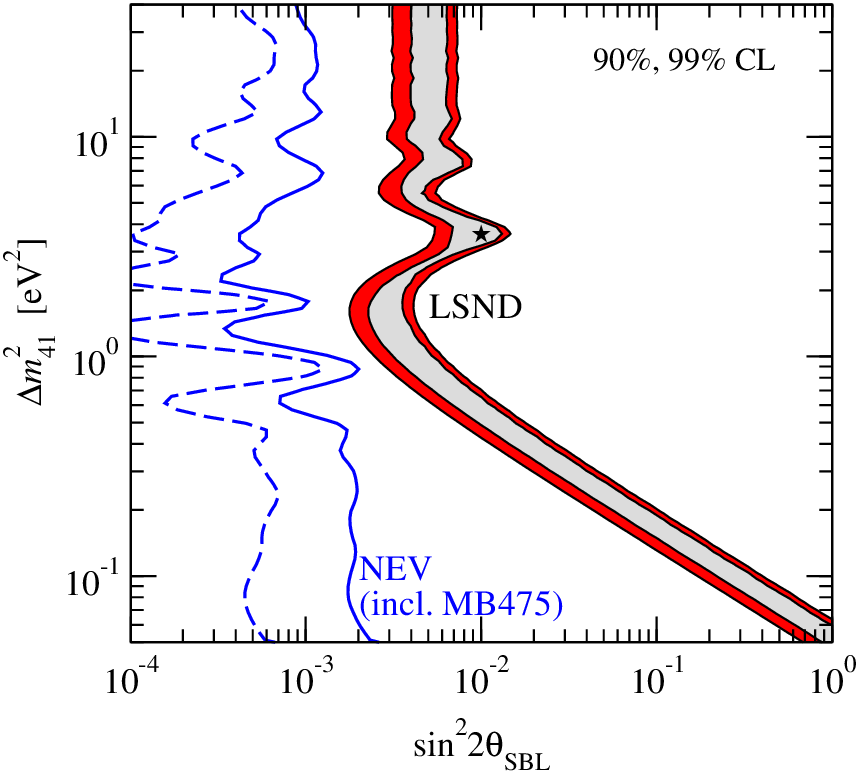}
  \caption{Left: Two-neutrino exclusion contours at 90\% C.L.\
  (2~d.o.f.) for MiniBooNE and KARMEN compared to the LSND allowed
  region at 90\% and 99\%~C.L. For all three experiments the same
  $\Delta\chi^2$ cut has been used to define the 90\%~C.L.\
  region. Right: Constraint on the LSND mixing angle in (3+1) schemes
  from no-evidence appearance and disappearance experiments (NEV) at
  90\% and 99\%~C.L. The shaded region corresponds to the allowed
  region from LSND decay-at-rest data.}
  \label{fig:LSND}
\end{figure}

Reconciling the LSND evidence~\cite{Aguilar:2001ty} for $\bar\nu_\mu
\to \bar\nu_e$ oscillations with the global neutrino data reporting
evidence and bounds on oscillations remains a long-standing problem
for neutrino phenomenology. Recently the MiniBooNE
experiment~\cite{polly,AguilarArevalo:2007it} added more information
to this question. This experiment searches for $\nu_\mu\to\nu_e$
appearance with a very similar $L/E_\nu$ range as LSND. No evidence
for flavour transitions is found in the energy range where a signal
from LSND oscillations is expected ($E > 475$~MeV), whereas an event
excess is observed below 475~MeV at a significance of
$3\sigma$. Two-flavour oscillations cannot account for such an excess
and currently the origin of this excess is under
investigation~\cite{polly}, see also~\cite{hill}.  MiniBooNE results
are inconsistent with a two-neutrino oscillation interpretation of
LSND at 98\%~CL~\cite{AguilarArevalo:2007it}, see
also~\cite{MBcomb}. The exclusion contour from MiniBooNE is shown in
Fig.~\ref{fig:LSND} (left) in comparison to the LSND allowed region
and the previous bound from the KARMEN experiment~\cite{karmen}, all
in the framework of 2-flavour oscillations.

%%%%%%%%%%%%%%%%%%%%%%%%%%%%%%%%%%%%%%%%%%%%%%%%%%%%%%%%%%%%%%%%%%
\section{Sterile neutrino oscillations}

The standard ``solution'' to the LSND problem is to introduce one or
more sterile neutrinos at the eV scale in order to provide the
required mass-squared difference to accommodate the LSND signal in
addition to ``solar'' and ``atmospheric'' oscillations.  However, in
such schemes there is sever tension between the LSND signal and
short-baseline disappearance experiments, most importantly
Bugey~\cite{Declais:1994su} and CDHS~\cite{Dydak:1983zq}, with some
contribution also from atmospheric neutrino
data~\cite{Bilenky:1999ny}.  I report here the results from a global
analysis including MiniBooNE data within schemes with one, two and
three sterile neutrinos~\cite{Maltoni:2007zf}.

\begin{figure}\
  \centering
  \includegraphics[height=6cm]{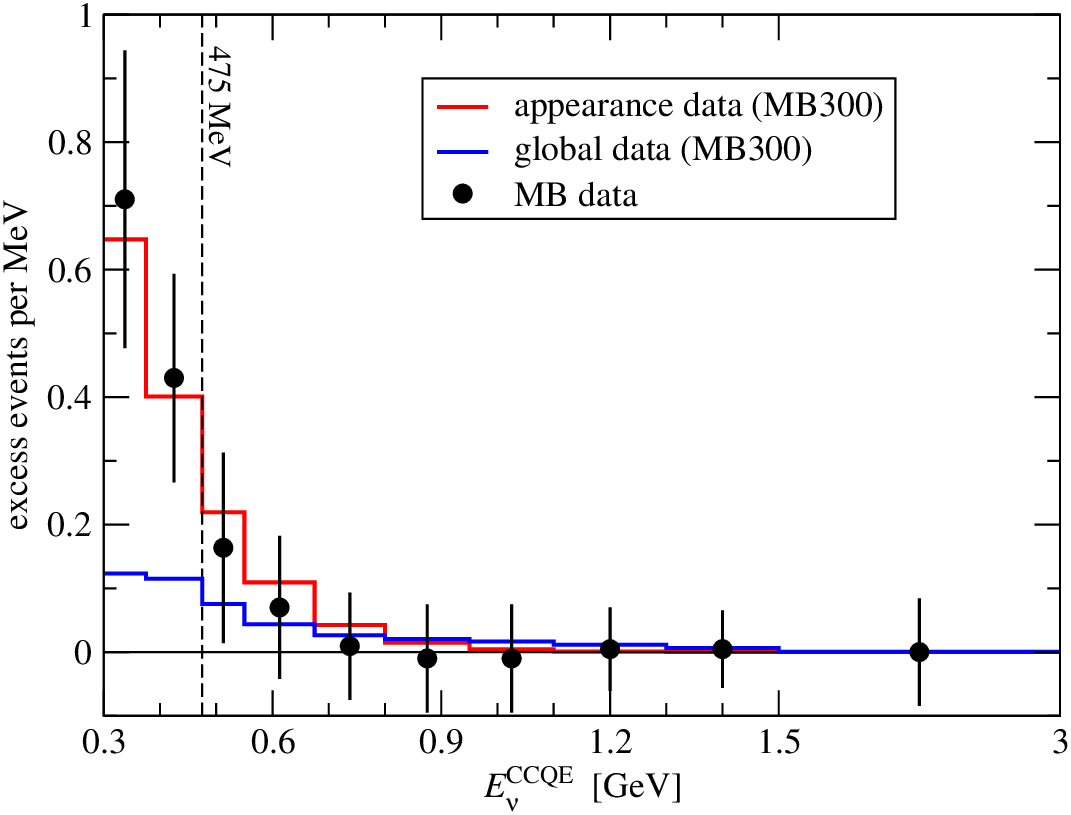}\quad
  \includegraphics[height=6cm]{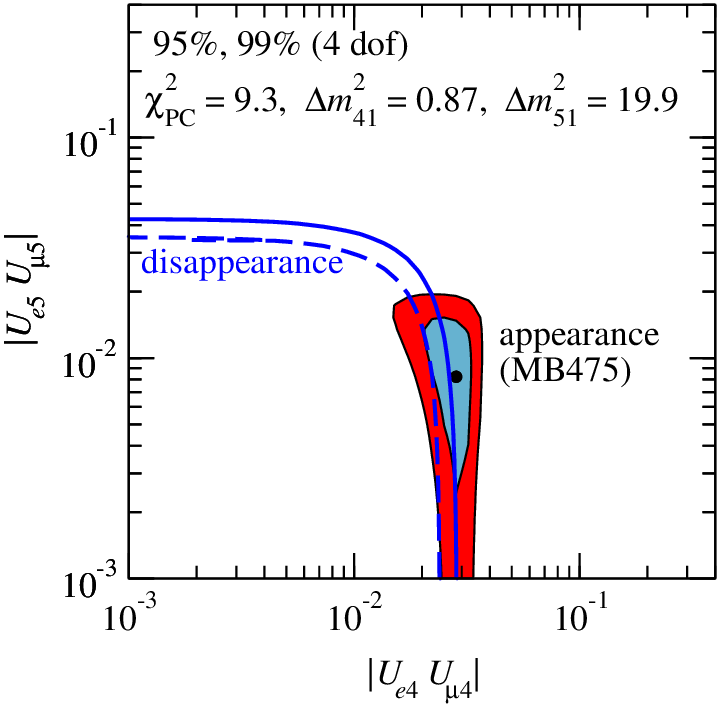}
  \caption{Left: Best fit spectra in (3+2) oscillations for MiniBooNE
  using appearance data only (MB, LSND, KARMEN, NOMAD) as well as in
  the global fit.  Right: Section of the 4-dimensional volumes allowed
  at 95\% and 99\%~CL in the (3+2) scheme from SBL appearance and
  disappearance experiments in the space of the parameters in common
  to these two data sets. The values of $\Dmq_{41}$ and $\Dmq_{51}$ of
  the displayed sections correspond to the point in parameter space
  where the two allowed regions touch each other (at a $\Delta\chi^2 =
  9.3$).}\label{fig:3+2}
\end{figure}

Four-neutrino oscillations within so-called (3+1) schemes have been
only marginally allowed before the recent MiniBooNE
results~\cite{Maltoni:2002xd,strumia,Sorel:2003hf}, and become even more
disfavored with the new data. We find that the LSND signal is
disfavoured by all other null-result short-baseline appearance and
disappearance experiments (including MiniBooNE) at the level of
$4\sigma$~\cite{Maltoni:2007zf}. The corresponding upper bound on the
effective LSND mixing angle is shown in Fig.~\ref{fig:LSND} (right).
Five-neutrino oscillations in (3+2) schemes~\cite{Sorel:2003hf} allow
for the possibility of CP violation in short-baseline
oscillations~\cite{Karagiorgi:2006jf}. Using the fact that in LSND the
signal is in anti-neutrinos, whereas present MiniBooNE data is based
on neutrinos, these two experiments become fully compatible in (3+2)
schemes~\cite{Maltoni:2007zf}. Moreover, in principle there is enough
freedom to obtain the low energy excess in MiniBooNE and being
consistent at the same time with the null-result in the high energy
part as well as with the LSND signal, see Fig.~\ref{fig:3+2} (left,
red histogram).
However, in the global analysis the tension between appearance and
disappearance experiments remains unexplained. This problem is
illustrated in Fig.~\ref{fig:3+2} (right) where sections through the
allowed regions in the parameter space for appearance and
disappearance experiments are shown. An opposite trend is clearly
visible: while appearance data require non-zero values for the mixing
of $\nu_e$ and $\nu_\mu$ with the eV-scale mass states 4 and 5 in
order to explain LSND, disappearance data provide an upper bound on
this mixing. The allowed regions touch each other at $\Delta\chi^2 =
9.3$, and a consistency test between these two data samples yields a
probability of only $0.18\%$, i.e., these models can be considered as
disfavoured at the $3\sigma$ level~\cite{Maltoni:2007zf}. Also,
because of the constraint from disappearance experiments the low
energy excess in MiniBooNE can not be explained in the global
analysis, see Fig.~\ref{fig:3+2} (left, blue histogram).
Furthermore, when moving from 4 neutrinos to 5
neutrinos the fit improves only by 6.1 units in $\chi^2$ by
introducing 4 more parameters, showing that in (3+2) schemes the
tension in the fit remains a sever problem. This is even true in the
case of three sterile neutrinos, since adding one more neutrino to
(3+2) cannot improve the situation~\cite{Maltoni:2007zf}.

%%%%%%%%%%%%%%%%%%%%%%%%%%%%%%%%%%%%%%%%%%%%%%%%%%%%%%%%%%%%%%%%%%%
\section{Exotic proposals}
\label{sec:exotic}

Triggered by these problems many ideas have been presented in order to
explain LSND, some of them involving very speculative physics, among
them sterile neutrino decay~\cite{Ma:1999im,Palomares-Ruiz:2005vf},
violation of the CPT~\cite{cpt,strumia,cpt-concha,cpt-4nu} and/or
Lorentz~\cite{lorentz} symmetries, quantum
decoherence~\cite{decoh-04,decoh-06,decoh} mass-varying
neutrinos~\cite{MaVaN}, short-cuts of sterile neutrinos in extra
dimensions~\cite{Pas:2005rb}, a non-standard energy dependence of
sterile neutrinos~\cite{Schwetz:2007cd}, or sterile neutrinos
interacting with a new gauge boson~\cite{Nelson:2007yq}. In the
following I comment on a personal selection of these exotic proposals,
without the ambition of being complete.

\bigskip\textit{CPT violation.}
Triggered by the observation that the LSND signal is in
anti-neutrinos, whereas their neutrino data is consistent with no
oscillations, it was proposed~\cite{cpt} that neutrinos and
anti-neutrinos have different masses and mixing angles, which violates
the CPT symmetry. A first challenge to this idea has been the KamLAND
reactor results, which require a $\Delta m^2$ at the solar scale for
anti-neutrinos. Subsequently it has been shown that the oscillation
signature in SuperK atmospheric neutrino data (which cannot
distinguish between $\nu$ and $\bar\nu$ events) is strong enough to
require a $\Delta m^2 \sim 2.5\cdot10^{-3}$~eV$^2$ for neutrinos as
well as for anti-neutrinos~\cite{cpt-concha},
see~\cite{GonzalezGarcia:2007ib} for an update. This rules out such an
explanation of the LSND signal with three neutrinos at
4.6$\sigma$. However, introducing a sterile neutrino, and allowing for
different masses and mixings for neutrinos and
anti-neutrinos~\cite{cpt-4nu} is fully consistent with all data,
including the MiniBooNE null-result in neutrinos. Such a model should
lead to a positive signal in the MiniBooNE anti-neutrino run.

\bigskip\textit{Sterile neutrino decay.}
Pre-MiniBooNE data can be fitted under the
hypothesis~\cite{Palomares-Ruiz:2005vf} of a sterile neutrino, which
is produced in pion and muon decays because of a small mixing with
muon neutrinos, $|U_{\mu 4}| \simeq 0.04$, and then decays into an
invisible scalar particle and a light neutrino, predominantly of the
electron type. One needs values of $g m_4\sim$~few~eV, $g$ being the
neutrino--scalar coupling and $m_4$ the heavy neutrino mass, e.g.\
$m_4$ in the range from 1~keV to 1~MeV and $g \sim
10^{-6}$--$10^{-3}$. This minimal model is in conflict with the
null-result of MiniBooNE. It is possible to save this idea by
introducing a second sterile neutrino, such that the two heavy
neutrinos are very degenerate in mass. If the mass difference is
comparable to the decay width, CP violation can be introduced in the
decay, and the null-result of MiniBooNE can be reconciled with the
LSND signal~\cite{Palomares-Ruiz:2005vf}.

\begin{figure}
  \centering
  \includegraphics[height=6.5cm]{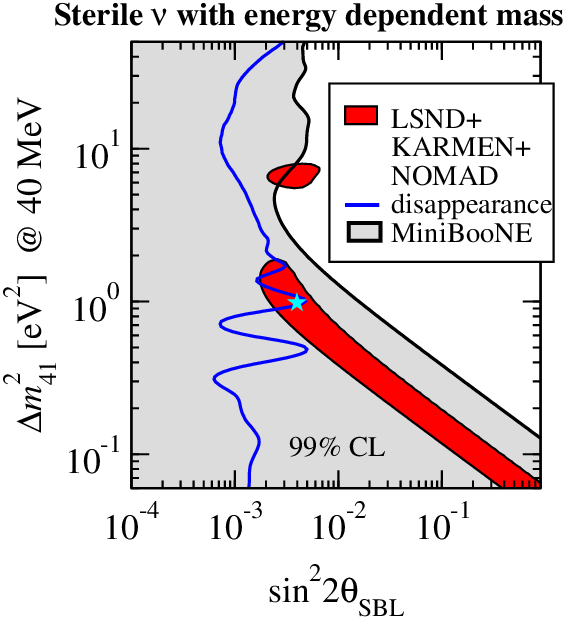}\qquad
  \includegraphics[height=6.5cm]{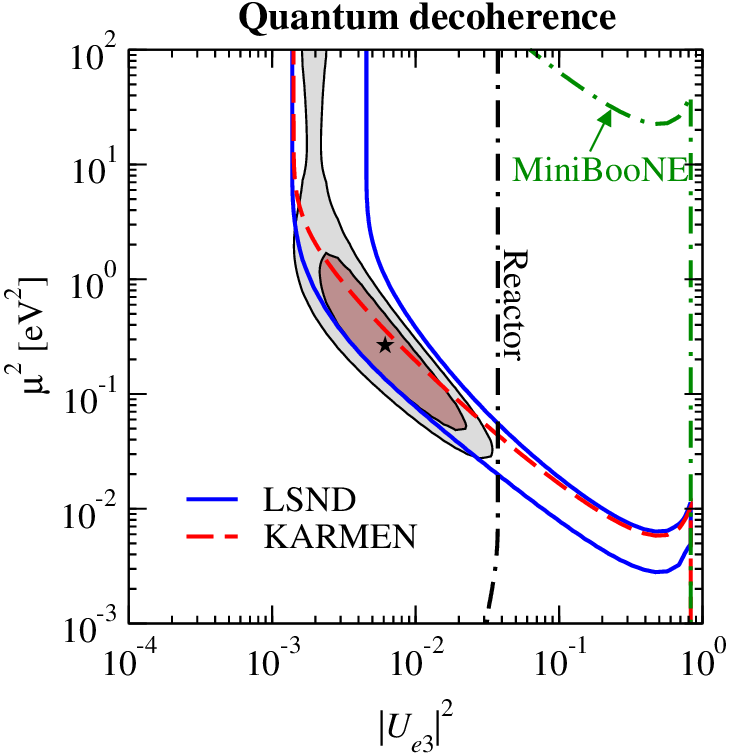}
  \caption{Left: Bounds from disappearance experiments and
  MiniBooNE compared to the LSND region for (3+1) oscillations when
  the sterile neutrino mass depends on energy as $m_4^2(E_\nu) \propto
  E_\nu^{-0.3}$. Right: Quantum decoherence in three-active neutrino
  oscillations. Lines correspond to 99\%~CL regions of individual
  experiments, shaded regions show the 90\% and 99\%~CL region of the
  global analysis, and the star marks the best fit point. The
  parameter $\mu$ is defined by parameterizing the decoherence
  parameter $\gamma$ as $\gamma = \mu^2/E_\nu \, (40\,{\rm
  MeV}/E_\nu)^3$.}\label{fig:exotic}
\end{figure}

\bigskip\textit{Sterile neutrinos with an exotic energy dependence.}
Short-baseline data can be divided into low-energy (few~MeV) reactor
experiments, LSND and KARMEN around 40~MeV, and the high-energy (GeV
range) experiments CDHS, MiniBooNE, NOMAD. Based on this observation
it turns out that the problems of the fit in (3+1) schemes can be
significantly alleviated if one assumes that the mass or the mixing of
the sterile neutrino depend on its energy in an exotic
way~\cite{Schwetz:2007cd}. For example, assuming that $m_4^2(E_\nu)
\propto E_\nu^{-r}$ one finds that for $r > 0$ the MiniBooNE exclusion
curve is shifted to larger values of $\Delta m^2$, whereas the bound
from disappearance experiments is moved towards larger values of the
mixing angle, and hence the various data sets become consistent with
LSND, compare Fig.~\ref{fig:exotic}~(left). At the best fit point with
$r \simeq 0.3$ the global fit improves by 12.7 units in $\chi^2$ with
respect to the standard (3+1) fit. Similar improvement can be obtained
if energy dependent mixing of the sterile neutrino is
assumed.\cite{Schwetz:2007cd}

Let us note that this is a purely phenomenological observation, and it
seems difficult to construct explicit models for such sterile
neutrinos. There are models which effectively introduce a non-standard
``matter effect'' for sterile neutrinos, {\it e.g.}\ via exotic extra
dimensions~\cite{Pas:2005rb} or via postulating a new gauge
interaction of the sterile neutrinos~\cite{Nelson:2007yq}. Similar as
in the usual MSW case, the sterile neutrino encounters effective mass
and mixing which depend on energy. However, in these approaches the
matter effect felt by the sterile state has to be some orders of
magnitude larger than the standard weak-force matter effect of active
neutrinos, in order to be relevant for short-baseline experiments.  In
such a case, in general very large effects are expected for
long-baseline experiments such as MINOS, atmospheric neutrinos, or
KamLAND. Unfortunately an explicit demonstration that a successful
description of all these data can be maintained in such models is still
lacking.

\bigskip\textit{Quantum decoherence.}
The possibility that the origin of the LSND signal might be quantum
decoherence in neutrino oscillations has been considered
in~\cite{decoh-04,decoh-06,decoh}.  Such effects can be induced by
interactions with a stochastic environment; a possible source for this
kind of effect might be quantum gravity.
The attempts to explain the LSND signal by quantum decoherence
in~\cite{decoh-04,decoh-06} seem to be in conflict with present
data. Both of these models are ruled out by the bound from NuTeV,
$P_{\nu_\mu \rightarrow \nu_e} , P_{\bar{\nu}_\mu \rightarrow
\bar{\nu}_e} < 5 \times 10^{-4}$ ($90\%$~C.L.)~\cite{NuTeV}.
Furthermore, the model of~\cite{decoh-04} (where in addition to
decoherence, CPT-violation is also introduced which results in a
difference between the oscillation probabilities for neutrinos and
anti-neutrinos) cannot account for the spectral distortion in the
anti-neutrino signal observed by KamLAND, whereas the scenario
of~\cite{decoh-06} is disfavored by the absence of a signal in KARMEN,
NOMAD and MiniBooNE.

Recently we have revisited this idea~\cite{decoh} by introducing a
different set of decoherence parameters. We assume that only the
neutrino mass state $\nu_3$ is affected by decoherence, whereas the
1-2 sector is completely unaffected, guaranteeing the standard
explanation of solar and KamLAND data. Hence, denoting as
$\gamma_{ij}$ the parameter which controls the decohering of the mass
states $\nu_i$ and $\nu_j$, we have $\gamma_{12} = 0$ and $\gamma_{13}
= \gamma_{23} \equiv \gamma$, where we have assumed that decoherence
effects are diagonal in the mass basis.
Furthermore, we assume that decoherence effects are suppressed for
increasing neutrino energies, $\propto E_\nu^{-r}$ with $r \sim
4$. This makes sure that at short-baseline experiments with $E_\nu
\gtrsim 1$~GeV such as MiniBooNE, CDHS, NOMAD, and NuTeV no signal is
predicted, and at the same time maintains standard oscillations for
atmospheric data and MINOS. In this way a satisfactory fit to the
global data is obtained. Disappearance and appearance data become
fully compatible with a probability of 74\%, compared to 0.2\% in the
case of (3+2) oscillations. The LSND signal is linked to the mixing
angle $\theta_{13}$, see Fig.~\ref{fig:exotic}(right) and hence, this
scenario can be tested at upcoming $\theta_{13}$ searches: while the
comparison of near and far detector measurements at reactors should
lead to a null-result because of strong damping at low energies, a
positive signal for $\theta_{13}$ is expected in long-baseline
accelerator experiments.

%%%%%%%%%%%%%%%%%%%%%%%%%%%%%%%%%%%%%%%%%%%%%%%%%%%%%%%%%%%%%%%%%%%%
\section{Outlook}

Currently MiniBooNE is taking data with anti-neutrinos.\cite{polly}
This measurement is of crucial importance to test scenarios involving
CP (such as (3+2) oscillations) or even CPT violation to reconcile
LSND and present MiniBooNE data. Therefore, despite the reduced flux
and detection cross section of anti-neutrinos the hope is that enough
data will be accumulated in order to achieve good sensitivity in the
anti-neutrino mode. Furthermore, it is of high importance to settle
the origin of the low energy excess in MiniBooNE. If this effect
persists and does not find an ``experimental'' explanation such as an
over-looked background, an explanation in terms of ``new physics''
seems to be extremely difficult. To the best of my knowledge, so-far no
convincing model able to account for the sharp rise with energy while
being consistent with global data has been provided yet.

The main goal of upcoming oscillation experiments like Double-Chooz,
Daya Bay, T2K, NO$\nu$A is the search for the mixing angle
$\theta_{13}$, with typical sensitivities of~\cite{Huber:2004ug}
$\stheta \gtrsim 1\%$. This should be compared to the size of the
appearance probability observed in LSND: $P_\mathrm{LSND} \approx
0.26\%$. Hence, if $\theta_{13}$ is large enough to be found in those
experiments sterile neutrinos may introduce some sub-leading effect,
but their presence cannot be confused with a non-zero $\theta_{13}$.
Nevertheless, I argue that it could be worth to look for sterile
neutrino effects in the next generation of experiments. They would
introduce (mostly energy averaged) effects, which could be visible as
disappearance signals in the near detectors of these experiments.
This has been discussed~\cite{Bandyopadhyay:2007rj} for the
Double-Chooz experiment, but also the near detectors at superbeam
experiments should be explored. An interesting effect of (3+2) schemes
has been pointed out recently for high energy atmospheric neutrinos in
neutrino telescopes~\cite{Choubey:2007ji}. The crucial observation is
that for $\Delta m^2 \sim 1$~eV$^2$ the MSW resonance occurs around
TeV energies, which leads to large effects for atmospheric neutrinos
in this energy range, potentially observable at neutrino telescopes.
Another method to test sterile neutrino oscillations would be to put a
radioactive source inside a detector with good spatial resolution,
which would allow to observe the oscillation pattern within the
detector~\cite{Grieb:2006mp}.
I stress that in a given exotic scenario such as the examples
discussed in sec.~\ref{sec:exotic} signatures in up-coming experiments
might be different than for ``conventional'' sterile neutrino
oscillations.

For the subsequent generation of oscillation experiments aiming at
sub-percent level precision to test CP violation and the neutrino mass
hierarchy, the question of LSND sterile neutrinos is highly
relevant~\cite{Donini:2001xp,Dighe:2007uf}. They will lead to a
miss-interpretation or (in the best case) to an inconsistency in the
results. If eV scale steriles exist with mixing relevant for LSND the
optimization in terms of baseline and $E_\nu$ of high precision
experiments has to be significantly changed. Therefore, I argue that
it is important to settle this question at high significance before
decisions on high precision oscillation facilities are taken.

\end{document}